\documentclass{iopart}
\usepackage{graphicx}  
\usepackage{latexsym}  

\jl{6}        % submitted to journal: Classical and Quantum Gravity
\eqnobysec    % section equation numbering

\def\beq{\begin{equation}}
\def\eeq{\end{equation}}
\def\rmd{{\rm d}}
\begin{document}

\title[Extended bodies in Kerr spacetime]
{Quadrupole effects on the motion of extended bodies in Kerr spacetime}

\author{
Donato Bini$^* {}^\S{}^\P$, Pierluigi Fortini$^\dagger$,
Andrea Geralico$^\ddag {}^\S$ and Antonello Ortolan$^{**}$}
\address{
  ${}^*$\
Istituto per le Applicazioni del Calcolo ``M. Picone,'' CNR I-00161 Rome, Italy
}
\address{
  ${}^\S$\
  ICRA, University of Rome ``La Sapienza,'' I-00185 Rome, Italy
  
}
\address{
${}^\P$
  INFN - Sezione di Firenze, Polo Scientifico, Via Sansone 1, 
  I-50019, Sesto Fiorentino (FI), Italy 
}
\address{
$^\dagger$
Department of Physics, University of Ferrara and INFN Sezione di Ferrara, I-44100 Ferrara,
Italy 
}

\address{
  ${}^\ddag$\
  Physics Department, University of Rome ``La Sapienza,'' I-00185 Rome, Italy
}

\address{
  ${}^{**}$\
  INFN - National Laboratories of Legnaro, I-35020 Legnaro (PD), Italy
}

\begin{abstract}
The motion of a body endowed with dipolar as well as quadrupolar structure is investigated in the Kerr background according to the Dixon's model, extending a previous analysis done in the Schwarzschild background. The full set of evolution equations is solved under the simplifying assumptions of constant frame components for both the spin and the quadrupole tensors and that the center of mass moves along an equatorial circular orbit, the total four-momentum of the body being aligned with it. We find that the motion deviates from the geodesic one due to the internal structure of the body, leading to measurable effects. 
Corrections to the geodesic value of the orbital period of 
a close binary system orbiting  
the Galactic Center are discussed assuming that the Galactic Center 
is a Kerr supermassive  black hole. 
\end{abstract}

\pacno{04.20.Cv}

\section{Introduction}

In a recent paper \cite{bfgo} we have investigated the motion of an extended body endowed with dipolar as well as quadrupolar structure in the field of a Schwarzschild black hole described by the Dixon's model \cite{dixon64,dixon69,dixon70,dixon73,dixon74}.
The corresponding evolution equations up to the quadrupole approximation are given by
\begin{eqnarray}\fl\qquad
\label{papcoreqs1}
\frac{DP^{\mu}}{\rmd \tau_U}&=&-\frac12R^{\mu}{}_{\nu\alpha\beta}U^{\nu}S^{\alpha\beta}-\frac16J^{\alpha\beta\gamma\delta}R_{\alpha\beta\gamma\delta}{}^{;\,\mu}
\equiv F^{\rm (spin)}{}^{\mu}+F^{\rm (quad)}{}^{\mu}\ , \\
\label{papcoreqs2}
\fl\qquad
\frac{DS^{\mu\nu}}{\rmd \tau_U}&=&2P^{[\mu}U^{\nu]}+\frac43J^{\alpha\beta\gamma[\mu}R^{\nu]}{}_{\gamma\alpha\beta}\ ,
\end{eqnarray}
where $P^{\mu}=m U_p^\mu$ (with $U_p\cdot U_p=-1$) is the total four-momentum of the particle, and $S^{\mu\nu}$ is a (antisymmetric) spin tensor; 
$U$ is the timelike unit tangent vector of the \lq\lq center of mass line'' ${\mathcal C}_U$ used to make the multipole reduction, parametrized by the proper time $\tau_U$.
The tensor $J^{\alpha\beta\gamma\delta}$ is the quadrupole moment of the stress-energy tensor of the body, and has the same algebraic symmetries as the Riemann tensor. 
We assume here that the only contribution to the complete quadrupole moment stems from the (symmetric) mass quadrupole moment $Q^{\alpha\beta}$ \cite{taub64,ehlers77}, implying that
\beq
J^{\alpha\beta\gamma\delta}=-3U_p^{[\alpha}Q^{\beta][\gamma}U_p^{\delta]}\ ,\qquad Q^{\alpha\beta}U_p{}_\beta=0\ .
\eeq

Moreover, the following additional conditions\cite{tulc59,dixon64} should be imposed to the spin tensor:
\beq
S^{\mu\nu}U_p{}_\nu=0\ ,
\eeq
to ensure the correct definition of the various multipolar terms.

It is convenient to introduce also the  spin vector by spatial (with respect to $U_p$) duality
\beq
\label{spinvec}
S^\beta={\textstyle\frac12} \eta_\alpha{}^{\beta\gamma\delta}U_p^\alpha S_{\gamma\delta}\ ,
\eeq
where $\eta_{\alpha\beta\gamma\delta}=\sqrt{-g} \epsilon_{\alpha\beta\gamma\delta}$ is the unit volume 4-form and $\epsilon_{\alpha\beta\gamma\delta}$ ($\epsilon_{0123}=1$) is the Levi-Civita alternating symbol, 
 as well as the scalar invariant
\beq
\label{sinv}
s^2=S^\mu S_\mu=\frac12 S_{\mu\nu}S^{\mu\nu}\ , 
\eeq
which, in general, is not a constant along the trajectory of the body. 

The compatibility of the model requires that the mass of the body, its spin as well as its quadrupole moments must all be small enough not to contribute significantly to the background metric. Otherwise, backreaction must be taken into account.

We have solved in \cite{bfgo} the system of equations (\ref{papcoreqs1})--(\ref{papcoreqs2}) under the simplifying assumptions of constant frame components (with respect to a natural orthonormal frame) of both the spin and the quadrupole tensor, obtaining the kinematical conditions to be imposed to the particle's structure in order the orbit of the particle itself be circular and confined on the equatorial plane of a Schwarzschild black hole.

We extend here this analysis to the more interesting case of a Kerr background maintaining the same restrictions on the  spin and quadrupole tensor components as well as on the circularity of motion in order to completely solve the problem, obtaining an analytical solution to be compared  with the corresponding one discussed in \cite{bfgo}.
As an example, we calculate the corrections to the geodesic value of the orbital period 
of a binary pulsar system, with the same parameters of PSR J0737-3039, orbiting close to
the Galactic Center black hole. 

Such analysis can also be extended to other  objects of astrophysical interest, 
e.g. ordinary or neutron stars, orbiting the Galactic Center 
supermassive black hole (Sgr A$^*$); in fact, recent measurements of near 
infrared periodic flares \cite{genzel,asche}  
suggest that Sgr A$^*$ is a rapidly rotating Kerr black hole with 
specific angular momentum in the range $(0.5 \div 1)M$. 
The interest to study orbits close to 
Sgr A$^*$ relies on the increasing accuracy in sub-milli-arcsecond astrometry by 
the near-infrared detectors \cite{eise} and on the potentiality of the 
next-generation radiotelescopes, e.g. the Square Kilometer Array (SKA) \cite{ska}, to identify 
some of the $10^4$ compact objects orbiting within 1 $parsec$ around 
Sgr A$^*$ \cite{muno}.

\section{Dynamics of extended bodies in the equatorial plane of a Kerr spacetime}

The Kerr metric  in standard Boyer-Lindquist coordinates is given by
\begin{eqnarray}
\rmd s^2 &=& -\left(1-\frac{2Mr}{\Sigma}\right)\rmd t^2 -\frac{4aMr}{\Sigma}\sin^2\theta\rmd t\rmd\phi+ \frac{\Sigma}{\Delta}\rmd r^2 +\Sigma\rmd \theta^2\nonumber\\
&&+\frac{(r^2+a^2)^2-\Delta a^2\sin^2\theta}{\Sigma}\sin^2 \theta \rmd \phi^2\ ,
\end{eqnarray}
where $\Delta=r^2-2Mr+a^2$ and $\Sigma=r^2+a^2\cos^2\theta$; here $a$ and $M$ are the specific angular momentum and total mass of the spacetime solution. The event horizons are located at $r_\pm=M\pm\sqrt{M^2-a^2}$. 

Let us introduce the Zero Angular Momentum Observer (ZAMO) family of fiducial observers, with four velocity
\beq
\label{n}
n=N^{-1}(\partial_t-N^{\phi}\partial_\phi);
\eeq
here $N=(-g^{tt})^{-1/2}$ and $N^{\phi}=g_{t\phi}/g_{\phi\phi}$ are the lapse and shift functions respectively. A suitable orthonormal frame adapted to  ZAMOs is given by
\beq
\label{zamoframe}
e_{\hat t}=n , \,\quad
e_{\hat r}=\frac1{\sqrt{g_{rr}}}\partial_r, \,\quad
e_{\hat \theta}=\frac1{\sqrt{g_{\theta \theta }}}\partial_\theta, \,\quad
e_{\hat \phi}=\frac1{\sqrt{g_{\phi \phi }}}\partial_\phi .
\eeq
We limit our analysis to  the equatorial plane ($\theta=\pi/2$) of the Kerr solution; as a convention, the physical (orthonormal) component along $-\partial_\theta$, perpendicular to the equatorial plane will be referred to as ``along the positive $z$-axis" and will be indicated by the index $\hat z$ when convenient: $e_{\hat z}=-e_{\hat \theta}$.
Moreover, 
we will consider our quadrupolar spinning test body as spinning only along the $\hat z$- direction and moving along  an equatorial circular orbit  of the Kerr spacetime, in the sense that both the center of mass line $U$ and the momentum $P=m U_p$ (with $U_p\cdot U_p=-1$) have only nonvanishing the $t$ and $\phi$ components:
\begin{eqnarray}
U&=&\gamma [e_{\hat t} +\nu e_{\hat \phi}], \qquad \gamma=(1-\nu^2)^{-1/2}\ ,\nonumber \\
U_p&=&\gamma_p\, [e_{\hat t}+\nu_p e_{\hat \phi}]\ , \quad \gamma_p=(1-\nu_p^2)^{-1/2}\ . 
\end{eqnarray}
We then proceed introducing observer-adapted frames to both $U_p$ and $U$ as follows.

An orthonormal frame adapted to $U_p$ is given by
\beq
\fl\qquad
e_0=U_p\ , \qquad e_1=e_{\hat r}\ , \qquad e_2=\gamma_p\, [\nu_p e_{\hat t}+ e_{\hat \phi}]\ , \qquad e_3=e_{\hat z}\ ,
\eeq
and hereafter all frame components of the various fields  are  meant to be referred to such a frame.
The spin vector results then 
$S=s\, e_3 $ 
and the orthogonality of the quadrupole tensor with respect to  $U_p$ implies
\beq
Q_{00}=Q_{01}=Q_{02}=Q_{03}=0\ .
\eeq
All the surviving components of the quadrupole tensor are constant along the path, corresponding to the definition of \lq\lq quasirigid motion"  due to Ehlers and Rudolph \cite{ehlers77}.  
Clearly in a more realistic situation the latter hypothesis should be released.

Being the evolution of the various quantities along $U$ 
it is also convenient to introduce 
a Frenet-Serret frame along $U$ \cite{iyer-vish}
\begin{eqnarray}\fl\quad
E_0\equiv U=\gamma [n+\nu e_{\hat \phi}]\ , \quad E_1=e_{\hat r}\ , \quad E_2\equiv E_{\hat \phi}=\gamma [\nu n+e_{\hat \phi}]\ , \quad  E_3=e_{\hat z}\ , 
\end{eqnarray} 
satisfying the following system of evolution equations
\begin{eqnarray}
\label{FSeqs}
\frac{DE_0}{d\tau_U}&=\kappa E_1\ , \qquad &  
\frac{DE_1}{d\tau_U}=\kappa E_0+\tau_1 E_2\ ,\nonumber \\
 \nonumber \\
\frac{DE_2}{d\tau_U}&=-\tau_1E_1+\tau_2E_3\ , \qquad &
\frac{DE_3}{d\tau_U}=-\tau_2E_2\ .
\end{eqnarray}

In this case, with $U$ tangent to an equatorial circular orbit, the second torsion $\tau_2$ is zero
while the geodesic curvature $\kappa$ and the first torsion $\tau_1$ are simply related by
\beq
\tau_1= -\frac{1}{2\gamma^2} \frac{\rmd \kappa}{\rmd \nu}\ .
\eeq
A direct calculation shows that
\beq
\label{ketau1}
\kappa =k_{\rm (lie)}\gamma^2 (\nu-\nu^g_+)(\nu-\nu^g_-)\ , 
\eeq
where $k_{\rm (lie)}$ is 
the Lie relative curvature of each orbit \cite{idcf2}
\beq
k_{\rm (lie)}\equiv -\partial_{\hat r} \ln \sqrt{g_{\phi\phi}}=-\frac{(r^3-a^2M)\sqrt{\Delta}}{r^2(r^3+a^2r+2a^2M)}\ ,
\eeq
and 
\beq
\nu^g_\pm =\frac{a^2\mp2a\sqrt{Mr}+r^2}{\sqrt{\Delta}(a\pm r\sqrt{r/M})}\ ,
\eeq 
are the linear velocities of co/counter-rotating geodesics on the equatorial plane of the Kerr spacetime.

Consider now first the evolution equations (\ref{papcoreqs2}) for the spin tensor. They imply that 
\beq
Q_{12}=Q_{13}=Q_{23}=0\ ,
\eeq
and introducing the \lq\lq structure functions" $f$ and $f'$ of the extended body, defined by
\beq
Q_{11}=Q_{33}+f\ , \qquad Q_{22}=Q_{33}+f'\ , 
\eeq
they also give
\begin{eqnarray}
\label{moto1}
0&=&[-\nu_p(\tau_1+\kappa\nu)+\nu\tau_1+\kappa]s-m(\nu-\nu_p)\nonumber\\
&&-f\frac{\gamma_p}{\gamma}[H_{\hat r \hat \theta}(1+\nu_p^2)-\nu_p(E_{\hat r \hat r}-E_{\hat \theta \hat \theta})]\ ,
\end{eqnarray}
where the electric and magnetic parts of the Weyl tensor with respect to the ZAMO frame (\ref{zamoframe}) have been introduced
\begin{eqnarray}
E_{\hat r \hat r} &=& -E_{\hat \theta \hat \theta}-E_{\hat \phi \hat \phi}\ , \nonumber\\
E_{\hat \theta \hat \theta}&=&\frac{M}{r^4}\frac{(r^2+a^2)^2+2a^2\Delta}{r^3+a^2r+2a^2M}\ ,\qquad  
E_{\hat \phi \hat \phi}= \frac{M}{r^3}\ ,  \nonumber\\
H_{\hat r \hat \theta} &=& -\frac{3Ma\sqrt{\Delta}}{r^4}\frac{r^2+a^2}{r^3+a^2r+2a^2M}\ .
\end{eqnarray}
By assuming the tracefree property characterizing the classical (euclidean) quadrupole moment tensor to hold also in the relativistic case, the components of quadrupole moment tensor $Q_{ab}$ in this case are completely determined by $f$ and $f'$
\begin{eqnarray}
\fl\qquad Q_{11}&=&\frac23 f-\frac13 f'\ , \quad Q_{22}=-\frac13f+\frac23 f'\ , \quad
Q_{33}=-\frac13 (f+f')\ .
\end{eqnarray}

Consider then the equations of motion (\ref{papcoreqs1}). 
They imply that
\begin{eqnarray}
\label{moto2old}\fl\quad
0&=&[H_{\hat r \hat \theta}(1+\nu\nu_p)-\nu_pE_{\hat r \hat r}+\nu E_{\hat \theta \hat \theta}]s -m[-\nu_p(\tau_1+\kappa\nu)+\nu\tau_1+\kappa]\nonumber\\
\fl\quad
&&+\frac{3}{2\gamma \gamma_p}\frac{M\sqrt{\Delta}}{r^5}\left\{
f'+f\,\gamma_p^2[c_1\nu_p+c_2(1+\nu_p^2)+c_3]
\right\}\ ,
\end{eqnarray}
where
\begin{eqnarray}
c_1&=&\frac{2a}{\sqrt{\Delta}}\frac{(r^2+a^2)^2+\Delta(4a^2+3r^2)}{r(r^3+a^2r+2a^2M)}\ , \nonumber\\
c_2&=&-\frac{(r^2+a^2)^2+4a^2(\Delta+Mr)}{r(r^3+a^2r+2a^2M)}\ , \nonumber\\
c_3&=&-\frac{(r^2+a^2)^2-a^2\Delta}{r(r^3+a^2r+2a^2M)}\ .
\end{eqnarray}

Solving Eqs. (\ref{moto1}) and (\ref{moto2old}) for $\nu$ and $\nu_p$ in terms of $s$ and $f$, $f'$ completely determines the motion.

According to \cite{bfgo} it is useful to introduce the following rescaled dimensionless angular and quadrupolar momentum quantities
\beq
\label{adim}
\sigma=\frac{s}{m}\zeta_K\ , \quad F=\frac{f}{m}\zeta_K^2\ , \quad F'=\frac{f'}{m}\zeta_K^2\ ,
\eeq
where $\zeta_K=(M/r^3)^{1/2}$ is the Keplerian value of the geodesic angular velocity in absence of background rotation (i.e. $a=0$), constant along $U$
due to the fact that along a circular orbit $r=\,$const. 

The quantities $\sigma$, $F$ and $F'$ are necessarily small. Although the quadrupolar terms $f$ and $f'$ are small only for a quasi-spherical body, the further rescaling by $\zeta_K$ makes indeed them small in any case. In fact, the radius of the orbit is assumed to be large enough in comparison with certain natural length scales like $|s|/m$ (also known as the M\o ller radius \cite{mol} of the body), $(|f|/m)^{1/2}$, $(|f'|/m)^{1/2}$ associated with the body itself in order to avoid backreaction effects.  

Eqs. (\ref{moto1}) and (\ref{moto2old}) then become
\begin{eqnarray}
\label{moto1_bis}
0&=&\zeta_K[-\nu_p(\tau_1+\kappa\nu)+\nu\tau_1+\kappa]\sigma-\zeta_K^2(\nu-\nu_p)\nonumber\\
&&-F\frac{\gamma_p}{\gamma}[H_{\hat r \hat \theta}(1+\nu_p^2)-\nu_p(E_{\hat r \hat r}-E_{\hat \theta \hat \theta})]\ ,
\end{eqnarray}
and
\begin{eqnarray}
\label{moto2_bis}
\fl\quad
0&=&\zeta_K[H_{\hat r \hat \theta}(1+\nu\nu_p)-\nu_pE_{\hat r \hat r}+\nu E_{\hat \theta \hat \theta}]\sigma -\zeta_K^2[-\nu_p(\tau_1+\kappa\nu)+\nu\tau_1+\kappa]\nonumber\\
\fl\quad
&&+\frac{3}{2\gamma \gamma_p}\frac{M\sqrt{\Delta}}{r^5}\left\{
F'+F\,\gamma_p^2[c_1\nu_p+c_2(1+\nu_p^2)+c_3]
\right\}\ ,
\end{eqnarray}
The above relations define the kinematical conditions allowing circular motion of the extended body taking into account its spinning and quadrupolar structures. 

Let us investigate the case of extended bodies with internal structure (dipolar and quadrupolar) compatible with a nearly geodesic motion. 
The case of a spinning particle with vanishing quadrupole moment tensor, i.e. $F=0=F'$, has been already studied in \cite{bdfgkerrclock}. 
In this situation Eqs. (\ref{moto1_bis})--(\ref{moto2_bis}) reduce to  
\begin{eqnarray}\fl\quad
0&=&[-\nu_p(\tau_1+\kappa\nu)+\nu\tau_1+\kappa]\sigma-\zeta_K(\nu-\nu_p)\ , \nonumber\\
\fl\quad
0&=&[H_{\hat r \hat \theta}(1+\nu\nu_p)-\nu_pE_{\hat r \hat r}+\nu E_{\hat \theta \hat \theta}]\sigma -\zeta_K[-\nu_p(\tau_1+\kappa\nu)+\nu\tau_1+\kappa]\ .
\end{eqnarray}
In the limit of small spin $\sigma$ we find
\begin{equation}
\label{nu_spin}
\nu_\pm=\nu^g_{\pm}+{\mathcal N}_\pm\sigma+O(\sigma^2)\ , \qquad
\nu_p^{(\pm)}=\nu_\pm+O(\sigma^2)\ ,
\end{equation}
where
\beq 
{\mathcal N}_\pm=\mp\frac{1}{2\zeta_K^2}[H_{\hat r \hat \theta}(1+\nu^g_{\pm}{}^2)-\nu^g_{\pm}(E_{\hat r \hat r}-E_{\hat \theta \hat \theta})]\ , 
\eeq
and the signs $\pm$ correspond to co/counter rotating orbits.
To first order in $a$ and neglecting also terms like $a\sigma$ the linear velocity (\ref{nu_spin}) becomes
\beq
\nu_\pm\simeq\pm\nu_K-3\nu_K\left[a\zeta_K+\frac{\sigma}{2}\right]\ , 
\eeq
where $\nu_K=[M/(r-2M)]^{1/2}$ is the Keplerian linear velocity.

If the contribution of quadrupolar terms can be considered negligible with respect to the dipolar ones and comparable with second order terms in the spin itself, one can consider corrections to Eq. (\ref{nu_spin}) as given by
\begin{eqnarray}
\fl\quad
\label{nunupsol}
\nu_\pm&\simeq&\nu^g_{\pm}+{\mathcal N}_\pm\sigma+{\mathcal N}_\pm\left\{
\sigma^2\left[1+\frac{{\mathcal N}_\pm}{\nu^g_+-\nu^g_-}\right]
-F\left[1-\frac{2H_{\hat r \hat \theta}\nu^g_{\pm}-E_{\hat r \hat r}+E_{\hat \theta \hat \theta}}{2\zeta_K^2}\right]
\right\}\ , \nonumber\\ 
\fl\quad
\nu_p^{(\pm)}&\simeq&\nu_{\pm} +2{\mathcal N}_\pm(F-\sigma^2)\ .
\end{eqnarray}
The corresponding angular velocity $\zeta_\pm$ and its reciprocal are
\begin{eqnarray}\fl\quad 
\label{zetasol}
\zeta_\pm&\simeq& \zeta^g_{\pm} +\frac{N{\mathcal N}_\pm}{\sqrt{g_{\phi\phi}}}\left\{
\sigma+\sigma^2\left[1+\frac{{\mathcal N}_\pm}{\nu^g_+-\nu^g_-}\right]
-F\left[1-\frac{2H_{\hat r \hat \theta}\nu^g_{\pm}-E_{\hat r \hat r}+E_{\hat \theta \hat \theta}}{2\zeta_K^2}\right]
\right\}\ , \nonumber\\ 
\fl\quad
\frac1{\zeta_\pm}&\simeq& \frac{1}{\zeta^g_{\pm}}(1+\lambda_d^\pm+\lambda_q^\pm)\ ,
\end{eqnarray}
where
\begin{eqnarray}
\lambda_d^\pm&=&-\frac{N{\mathcal N}_\pm}{\sqrt{g_{\phi\phi}}\zeta^g_{\pm}}\,\sigma\ , \nonumber\\
\lambda_q^\pm&=&-\frac{N{\mathcal N}_\pm}{\sqrt{g_{\phi\phi}}\zeta^g_{\pm}}\left\{\sigma^2\left[1+{\mathcal N}_\pm\left(\frac{1}{\nu^g_+-\nu^g_-}-\frac{N}{\sqrt{g_{\phi\phi}}\zeta^g_{\pm}{}^2}\right)\right]\right.\nonumber\\ 
&&\left.-F\left[1-\frac{2H_{\hat r \hat \theta}\nu^g_{\pm}-E_{\hat r \hat r}+E_{\hat \theta \hat \theta}}{2\zeta_K^2}\right]
\right\}\ .
\end{eqnarray}
The period of revolution around the central source thus turns out to be 
\beq
\label{periodo2}
T=\frac{2\pi}{|\zeta_\pm|}=T^g_\pm\left|1+\lambda_d^\pm+\lambda_q^\pm\right|\ , \qquad
T^g_\pm=\frac{2\pi}{|\zeta^g_\pm|}\ .
\eeq  

In the limit of small values of the black hole rotation parameter $a$ Eqs. (\ref{nunupsol})--(\ref{zetasol}) reduce to 
\begin{eqnarray}
\nu_\pm&\simeq&\pm\nu_K-3\nu_K\left[a\zeta_K+\frac{\sigma}{2}\right]\pm\frac32\frac{\zeta_K}{\nu_K}(1+4\nu_K^2)a\sigma\pm\frac38\nu_K(\sigma^2+2F)\nonumber\\
&&-\frac34\frac{\zeta_K}{\nu_K}a\left[(1+10\nu_K^2)F-(1+4\nu_K^2)\sigma^2\right]\ , \nonumber\\
\nu_p^{(\pm)}&\simeq&\nu_{\pm} \pm 3\nu_K(F-\sigma^2)-3\frac{\zeta_K}{\nu_K}(1+4\nu_K^2)a(F-\sigma^2)\ , \nonumber\\
\zeta_\pm&\simeq&\frac{\zeta_K}{\nu_K}\nu_\pm\ , \nonumber\\
\frac1{\zeta_\pm}&\simeq&\pm\frac1{\zeta_K}+a+\frac{3}{2\zeta_K}\sigma\mp\frac32\frac{r}{M}a\sigma\mp\frac{3}{8\zeta_K}(2F-5\sigma^2)\nonumber\\
&&+\frac32a\left[3(F-\sigma^2)+\frac{r}{2M}(F-7\sigma^2)\right]\ ,
\end{eqnarray}
from which the limiting expression for the period of revolution $T=2\pi/|\zeta_\pm|$ follows easily. 

A direct measurement of $T$ will then allow to estimate the quantity $F$ determining the quadrupolar structure of the body, if its spin is known.
Note that the fraction $\lambda_d^\pm$ due to the spin is different depending on whether the body is spinning up or down, whereas the term $\lambda_q^\pm$ due to the quadrupole has a definite sign once the shape of the body is known ($F$ cannot change its sign).

\section{Applications}

An interesting opportunity to test the quadrupole effect of extended body would arise for instance from the motion of ordinary or neutron stars around Sgr A$^*$, the supermassive ($M\simeq 10^6\ M_\odot$) rotating ($a\in [0.5, 1] M$) black hole located at the Galactic Center \cite{falcke,genzel,muno}. 

To illustrate the order of magnitude of the effect, we may consider a binary pulsar system 
similar to PSR J0737-3039 as orbiting Sgr A$^*$ at a distance of $r\simeq 10^9$ Km.  
The PSR J0737-3039 system consists of two close neutron stars (their separation is only $d_{AB} \sim 8 \times 10^5$ Km) of comparable masses $m_A\simeq 1.4\ M_\odot$, $m_B \simeq 1.2\ M_\odot$), but very different intrinsic spin period ($23$ ms of pulsar A vs $2.8$ s of pulsar B) \cite{lyne}. 
Its orbital period is about $2.4$ hours, the smallest yet known for such an object. Since the intrinsic rotations are negligible with respect to the orbital period, we can treat the binary system as a single object with reduced mass $\mu_{AB} \simeq 0.7 \ M_\odot$ and intrinsic rotation equal to the orbital period. 
The spin parameter thus turns out to be equal to $\sigma\approx6\times10^{-8}$, whereas the quadrupolar parameters are $F=F'\approx9.6\times10^{-10}$, since we have taken $f=f'=\mu_{AB}d_{AB}^2$ as a rough estimate.

Since the rotation parameter of Sgr A$^*$ is not small we must use the exact expression (\ref{periodo2}).
In the literature one finds different estimates of the spin parameter of the Galactic Center black hole, ranging from 
$a\simeq 0.52 M$ \cite{ghez} all the way up to  $a\simeq 0.983 M$ \cite{stuc1,stuc2} or even $a\simeq 0.996 M$ \cite{aschen04,aschen07}. 
We list in Table \ref{tab:1} the corresponding
values of the geodesic period $T^g_+$ of the PSR J0737-3039 bynary system as well as the corrections $\lambda_d^\pm$ and $\lambda_q^\pm$ due to its dipolar and quadrupolar structure respectively.

% table 1

\begin{table}
\begin{center}
\begin{tabular}{|c|c|c|c|}
\hline
\rule{0pt}{3ex}$a/M$ & $T^g_+ ({\rm cm})$ & $\lambda_d^+$ & $\lambda_q^+$\\
\hline
\rule{0pt}{2ex} $0.52$ & $1.62236\times10^{16}$ & $8.83\times10^{-8}$ & $-7.05\times10^{-10}$\\
\rule{0pt}{2ex} $0.75$ & $1.62238\times10^{16}$ & $8.75\times10^{-8}$ & $-6.99\times10^{-10}$\\
\rule{0pt}{2ex} $0.983$ & $1.62240\times10^{16}$ & $8.66\times10^{-8}$ & $-6.92\times10^{-10}$\\
\rule{0pt}{2ex} $0.996$ & $1.62241\times10^{16}$ & $8.66\times10^{-8}$ & $-6.92\times10^{-10}$\\
\rule{0pt}{2ex} $1$ & $1.62241\times10^{16}$ & $8.66\times10^{-8}$ & $-6.92\times10^{-10}$\\
\hline
\end{tabular}
\end{center}
\caption{The estimates of geodesic period $T^g_+$ as well as the corrections $\lambda_d^+$ and $\lambda_q^+$ due to both the dipolar and quadrupolar structure of the PSR J0737-3039 bynary system are listed for different values of Sgr A$^*$ rotational parameter $a/M$.
Note that in order to resolve the dipolar and quadrupolar effects the measured period should be known with very high accuracy.}
\label{tab:1}
\end{table}

It turns out that no relevant differences arise for the selected values of the black hole spin parameter.
In particular, the results are not sensitive of the black hole being either fast rotating \cite{stuc2} or near extreme \cite{aschen07} or even extreme ($a=M$). 
In the latter case Eqs. (\ref{periodo2}) reduce to
\begin{eqnarray}
\label{extreme_case}
\fl\quad
T^g_\pm&=&2\pi  \left(M\pm \frac{1}{\zeta_K}\right)\ , \nonumber\\
\fl\quad
\lambda_d^\pm &=& \pm \frac32 \sigma \Lambda^\pm\ , \nonumber\\
\fl\quad
\lambda_q^\pm &=& - \frac34 \Lambda^\pm \left[
\pm F \left( 1-6r^3\zeta_K^3 \Lambda^+ \right) +2\sigma^2 \left( \pm 1 -\frac34 
\Lambda^\pm (2\pm 1-r^3\zeta_K^3)
\right)
\right]\ ,
\end{eqnarray}
where
\beq
\Lambda^\pm = \frac{1-r\zeta_K}{1+r^3\zeta_K^3} \frac{1\pm r^3\zeta_K^3}{1+r^3\zeta_K^3}\ .  
\eeq

The detection of pulsars in Sgr A$^*$ is difficult because of the intense scattering region 
located in front of Sgr A$^*$. However,  these pulsars may be detectable in the next decade
by the SKA detector, which promises 
high frequency  sensitivity and large collecting area \cite{ska}. 

Another possible application of our calculations could be the orbital motions of the 
so called S-stars \cite{ghez}, i.e. the massive ($(30\div 120)\ M_\odot$ ), 
young ($<10\ Myr$) stars within the influence 
of the supermassive black hole. However, in this case the orbits are no longer circular 
and the problem of discriminating quadrupole effects would deserve 
further investigation.

\section{Concluding remarks}

We have investigated the motion of extended bodies endowed with dipolar as well as quadrupolar structure on a Kerr background following Dixon's model, generalizing previous results. 
Under the simplified assumptions of constant frame components (with respect to a natural orthonormal frame) of both the spin and the quadrupole tensor describing the body we have found the kinematical conditions to be imposed to the particle's structure in order the orbit of the particle itself be circular and confined on the equatorial plane.
The motion turns out to deviate from the geodesic one due to the internal structure of the body, leading to measurable effects. 
The effect of the quadrupole terms could be important for instance when considering the period of revolution of an extended body around the central source.
We have then applied such analysis to the binary pulsar system PSR J0737-3039 as if it were orbiting the Galactic Center black hole, providing an estimate of the corrections to the geodesic value of the orbital period due to the dipolar as well as quadrupolar terms.

\end{document}